\documentclass[nologo,nourl,11pt,a4paper]{ETHpaper}

\usepackage{verbatim}  

\graphicspath{{figures/}}

\usepackage{color}

\renewcommand*{\=}{{\kern0.1em=\kern0.1em}}
\renewcommand*{\-}{{\kern0.1em-\kern0.1em}} 
\newcommand*{\+}{{\kern0.1em+\kern0.1em}}

\title{Fragile, yet resilient: Adaptive decline \\ in a collaboration network of firms}
\author{Frank Schweitzer,$^{1,2,*}$  Giona Casiraghi,$^{1}$ Mario V. Tomasello,$^{1}$ David Garcia$^{2,3}$ }
\address{\small 
  $^1$Chair of Systems Design, ETH Zurich, Switzerland \\
  $^{2}$Complexity Science Hub Vienna, Austria \\
  $^{3}$Institute of Interactive Systems and Data Science, Graz University of Technology,
  Austria \\
$^\ast$Corresponding author; E-mail: fschweitzer@ethz.ch\\

}
\reference{\rm (Submitted for publication)} 
\titlealternative{Fragile, yet resilient: Adaptive decline in a collaboration network of firms}
\authoralternative{F. Schweitzer, G. Casiraghi, M. V. Tomasello, D. Garcia}

\begin{document}

\maketitle

\begin{abstract}
  \noindent
  The dynamics of collaboration networks of firms follow a life-cycle of growth and decline. 
  That does not imply they also become less resilient.
  Instead, declining collaboration networks may still have the ability to mitigate shocks from firms leaving, and to recover from these losses by adapting to new partners.
  To demonstrate this, we analyze 21.500 R\&D collaborations of 14.500 firms in six different industrial sectors over 25 years.
  We calculate time-dependent probabilities of firms leaving the network and simulate drop-out cascades, to determine the expected dynamics of decline.
  We then show that deviations from these expectations result from the adaptivity of the network, which mitigates the decline.
These deviations can be used as a measure of network resilience. 
\end{abstract}

\section{Introduction}
\label{sec:introduction}

\emph{Resilience} denotes the ability of a system to \emph{withstand} shocks and to \emph{recover} from them~\citep{Hollnagel2007,Sterbenz2010}.
Hence, it combines two different dimensions: \emph{robustness} against shocks and \emph{adaptivity} to overcome states that result from a shock~\citep{Walker2004,Johnson2010a}.
Interestingly, most research has only focused on the first aspect, robustness.
Much less attention is paid to the second one, which is more difficult to quantify and to forecast.
Therefore, in this paper we aim at better understanding the adaptive capacity of systems.

One reason for the biased research interest comes from the fact that robustness is strongly related to concepts like \emph{stability} which are easier to assess.
However, if robustness or stability are used as synonyms for resilience~\citep{Holling1973}, 
the temporal aspects of \emph{recovery} are neglected~\citep{Kitano2004,Scheffer2012}. 
To quote Abraham Lincoln: ``It's not important how many times you fall, but how many times you get back up''. 
If we want to improve the resilience of systems, the solution is not to simply avoid situations that may lead to a breakdown, by increasing the robustness of a system \cite{Interventions2020}.
Very often such breakdowns cannot be avoived or even controlled \cite{ZhangACS2019,Firms2020}.  
The real problem is how to enable systems to cope with these situations and to recover from them~\citep{sutcliffe2003organizing,Lengnick2005}.

This requires us to develop a systemic view that takes the \emph{eigendynamics} of a system into account~\citep{Mcmanus2007,Goggins2014}.
To address this, we need an appropriate system representation.
The \emph{complex systems} perspective assumes a large number of interacting system elements, denoted as agents.
Such systems can be visualized as \emph{complex networks} in which agents are represented by nodes and their interactions by links.
In our paper, we adopt this perspective to model \emph{collaboration networks} in economics, i.e., nodes represent firms and links their joint activities in research and development (R\&D) \cite{Battiston2009,ACS-econ-netw-2009}.

Similar to engineered systems, 
many social, economic and biological systems follow a \emph{life cycle}~\citep{gulati2012rise,powell2005network}.
After an initial growth phase one observes a period of maturity or saturation, which eventually leads to the decline and the decommision of the system \cite{Saavedra_2008}. 
Also collaboration networks between firms follow such a life cycle \cite{ICC2016rise}.
Maintaining collaborations is costly, but only through collaborations firms have access to knowledge they do not develop in-house \cite{JEE2018}.
Hence, firms will collaborate as long as they obtain a benefit from this.
If the goal of the collaboration, e.g., patent development or knowledge exchange, is fulfilled they will delete the link to the respective partner. 
From the \emph{life cycle} perspective, the decay of the collaboration network is therefore not a sign of weakness or malfunction, but a sign of a quasi-natural, even rational dynamics.

If one supposes that network \emph{growth} indicates a \emph{positive}  and network \emph{decline} a \emph{negative} development, this would imply that shrinking networks are less resilient.
With our study, we want to challenge such a na\"ive argumentation. 
The title phrase ``fragile, yet resilient'' summarizes our main finding that the system at hand even in phases of decline has the ability to respond to this development, by \emph{adaptation}.
This does not mean that the collaboration system completely recovers.
But it is interesting to note that the decline can be stopped and the loss can be mitigated.
Such a development is most often overlooked, simply because the observed dynamics is dominated by the global (negative) trend.
Therefore, our aim is to detect the adaptive capability of the system, and to separate it from the global trend, this way quantifying the recovery potential.

\section{The collaboration network of firms}
\label{sec:coll-netw-firms}

\subsection{Data and networks}
\label{sec:data}

\begin{figure}[htbp]
 \includegraphics[width=0.45\textwidth]{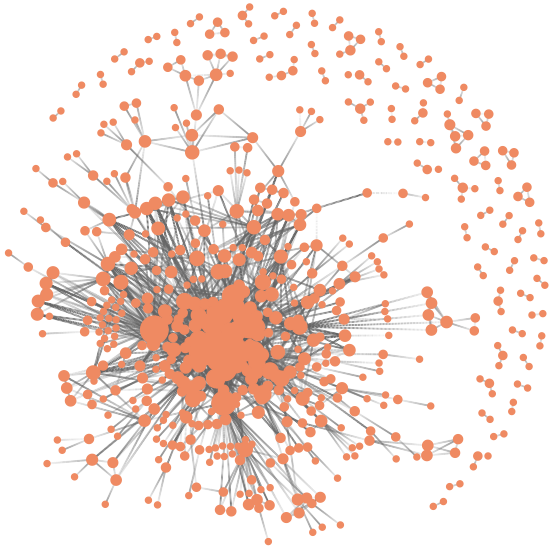}
 \hfill
 \includegraphics[width=0.45\textwidth]{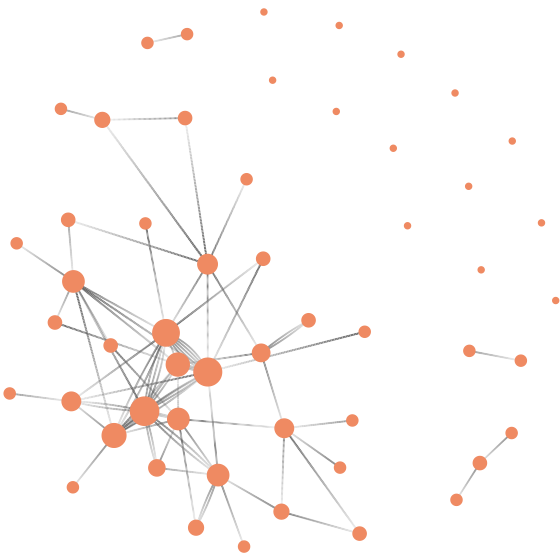} \\

\caption{Network of R\&D collaborations of firms in the sector ``Communications equipment'' in 1995 (left). The right plot shows how many of the firms from 1995 are still present in 2003. }
  \label{fig:network}
\end{figure}

Firms with a focus on research and development (R\&D) activities continuously establish new collaborations with other firms, to exchange knowledge and to leverage synergies. 
Because firms have to declare their R\&D alliances, we have access to a large data set of more than 14.500 firms and 21.500 collaborations over a time interval of 25 years (1984-2009), covering six different industrial sections, e.g., \emph{Pharmaceuticals} or \emph{Computer Hardware}.
The details of the data set are described in our different publications \cite{Perra2014,ICC2016rise,EPJDS-Cross-2017a}. 
Figure~\ref{fig:network} shows an example of such a collaboration network in two different years. 

An empirical investigation of the evolution of these sectoral collaboration networks has  revealed a \emph{rise and fall dynamics}, also illustrated in Figure~\ref{fig:network}. 
Two periods in the evolution of these collaboration networks have been distinguished: From 1984 to 1995, we see a steady growth of the networks, both with respect to the number of firms and the number of links.
From 1995 to 2009, on the other hand, we observe that these networks continuously shrink. 
This holds despite the mentioned fact that firms continue to establish new alliances.
But theses activities do not break the declining trend.

\subsection{Leaving probability}
\label{sec:leaving-probability}

In the following, we seek to quantify the tendency of firms to leave. 
Our considerations start from the question why firms stay in a collaboration network.
As other economic actors, firms try to maximize their utility, i.e., the difference between benefits and costs.
Hence, firms stay  as long as their benefits exceed their costs.
But even if firms leave, they can still return to the network later to start new R\&D collaborations with the same or with other partners.

While this dynamics seems reasonable, we have to overcome the problem that there is only data available about the \emph{starting date} when firms establish a new alliance, but no data about the \emph{ending date}. 
Thus, we first need to estimate the life time of an R\&D alliance.
This problem was solved in a subsequent study \citep{ACS2016}.
We have estimated that the mean life time of an R\&D alliance is about 3 years.
We build on this result here, assuming that the life time of an alliance is randomly drawn from a normal distribution with a mean of 3 years and a standard deviation of 1 year. 

This life time estimation has enabled us to reconstruct the evolution of the collaboration network as detailed in \citep{ICC2016rise,Perra2014}.
We use the starting date of each alliance and the information about its collaboration partners.
Then, we sample a life time of the alliance from the mentioned distribution to determine its ending date, at which we \emph{remove} all collaboration \emph{links} related to that alliance.
The end of an alliance does not imply that firms leave the collaboration network.
In the meantime they may have used their presence to establish new alliances with other firms.
Only if firms have \emph{no active} alliances in a given year, they will leave.
This information is aggregated for each year $t$. 

Once we know which firms stayed and which firms left, we calculate the leaving probability $p$ as follows. 
For each firm, we use a time dependent state variable, $s_{i}(t)=1$ if firm $i$ is present in year $t$ and $s_{i}(t)=0$ if it is absent. 
The probability to \emph{leave} is then defined as:
$p_{i}(t)=p[s_{i}(t+1)=0|s_{i}(t)=1]$, i.e., it is the probability that a firm present in year $t$ is absent in the following year.
The probability to \emph{stay} is $1-p_{i}(t)$.

\begin{figure}[htbp]
\centering
\includegraphics[width=.3\textwidth]{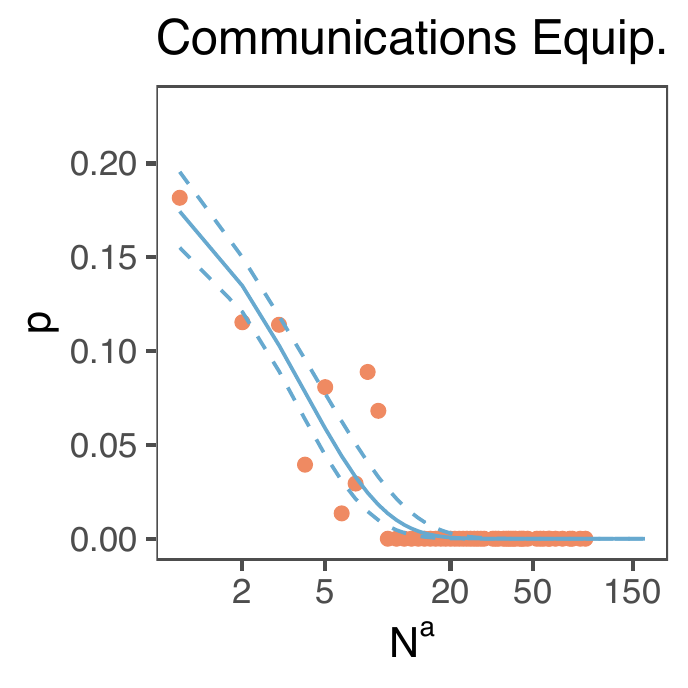}\hfill
\includegraphics[width=.3\textwidth]{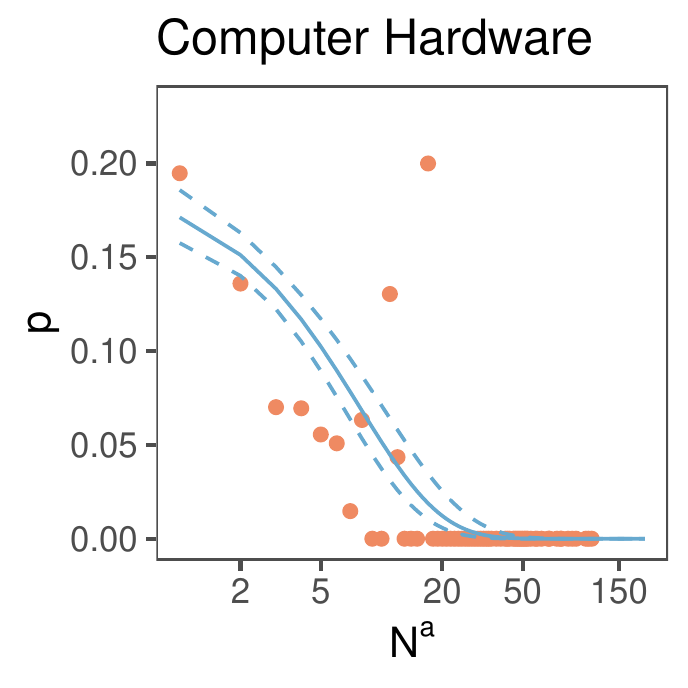}\hfill
\includegraphics[width=.3\textwidth]{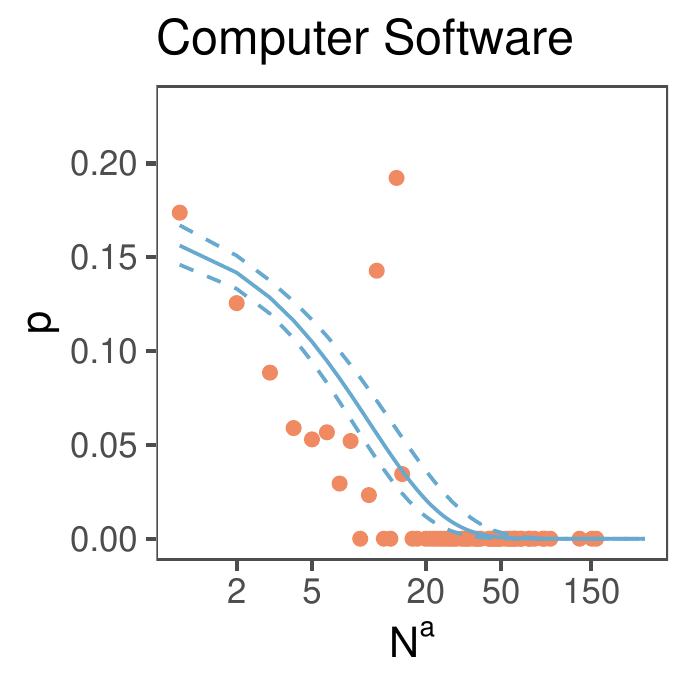}\\
\includegraphics[width=.3\textwidth]{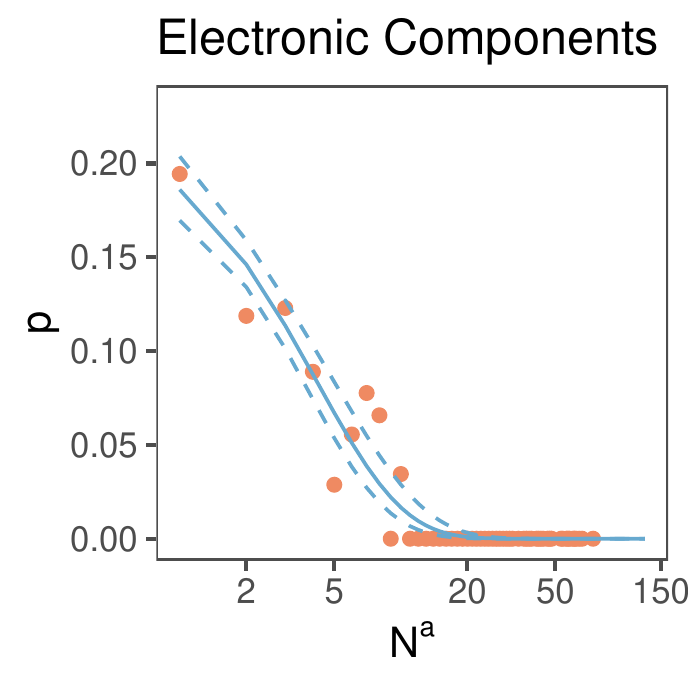}\hfill
\includegraphics[width=.3\textwidth]{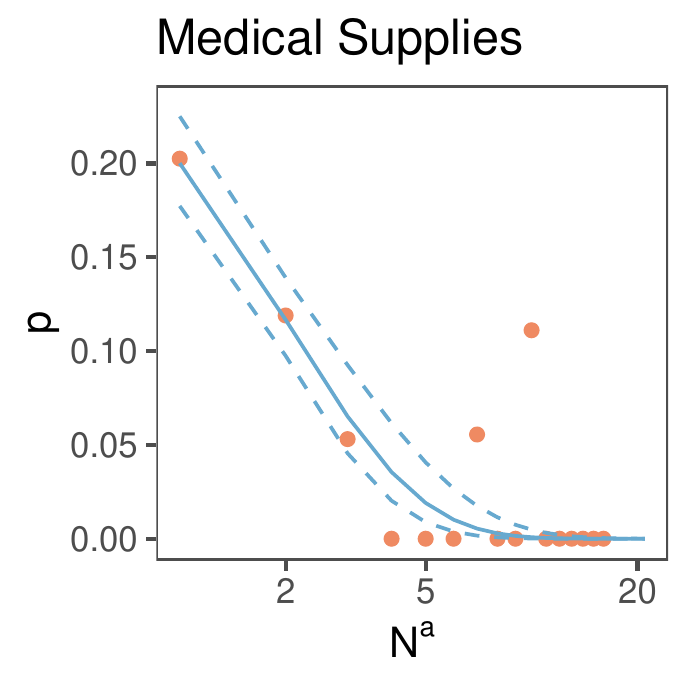}\hfill
\includegraphics[width=.3\textwidth]{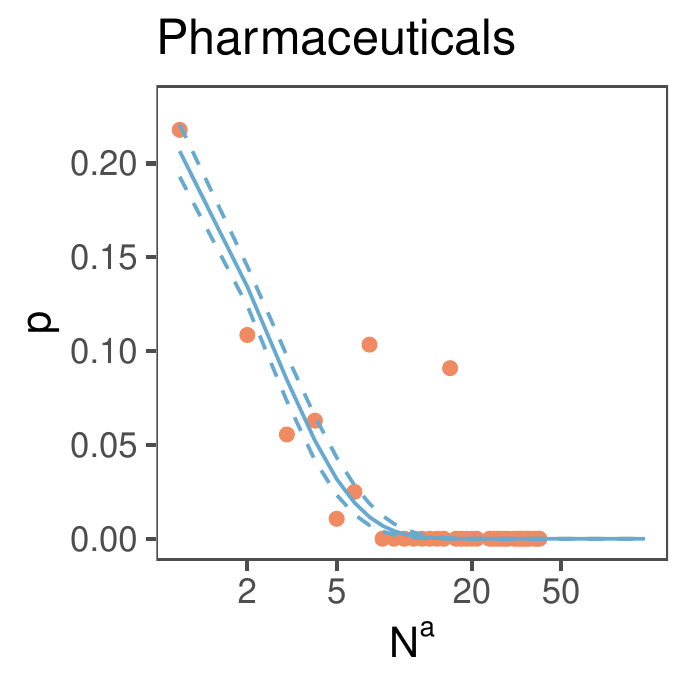}
\caption[]{Probability of firms to leave the network as a function of the number of their active partners. 
}
\label{fig:leaving_probabilities}
\end{figure}

Our aim in this paper is to estimate how the leaving probability $p_{i}$ depends on the benefits of a firm. 
We argue that, given the aim of the collaboration is knowledge exchange, the benefits of firm $i$ crucially depend on its number of active partners $N^{a}_{i}$ in the collaboration network.
We conjecture, the better the firm's embeddedness in the network, i.e., the more active partners, the less the probability to leave.
To obtain a quantitative relation, we first measure, for each year $t$, the number of active partners $N^{a}_{i}(t)$ of each firm present in the network.
Then, we determine for the same year its leaving probability $p_{i}(t)$ as introduced above and define the relation to the number of active partners as:
\begin{align}
  p_{i}(t)=p[s_{i}(t+1)=0|s_{i}(t)=1] \propto \exp\left\{\alpha+\beta N^{a}_{i}(t)\right\}
  \label{eq:2}
\end{align}
As a reference, we first want to estimate $p$ for the period of network \emph{growth} ending in 1995. 
Therefore we aggregate all data for the period from 1984-1995, and then do a logistic regression on the log odds (or logit),
$\ln\left[{p}/{(1-p)}\right]=\alpha+\beta N^{a}$.
The regression results are shown in Figure~\ref{fig:leaving_probabilities} together with the empirical leaving probabilities (yellow marks) for the six different industrial sectors.

We clearly see the monotonous decrease of the leaving probabilities with the number of active partners.
The plotted 95\% confidence intervals of the estimates indicate that the results are indeed reliable.
At the same time, we also notice the differences between industrial sectors, in particular regarding the number of active partners.

\section{Dynamic modeling of cascades}
\label{sec:dynam-model-casc}

\subsection{Time dependent leaving probability}
\label{sec:const-leav-prob}

Our task is now to model the cascades of firms leaving the collaboration network.
We start from the network at its maximum size in 1995 and only consider firms that are present there in 1995, which we call the ``class of 95'' in the following. 
If the total number of firms in the network is $N^{\mathrm{tot}}(t)$ and the number of firms remaining from the class of 95 is $N(t)$, by construction, in our reference year 1995 $N=N^{\mathrm{tot}}$.
Afterwards, $N(t)$ will decrease because of firms leaving, and the question is how fast this decline happens.

\begin{figure}[htbp]
\centering
\includegraphics[width=.3\textwidth]{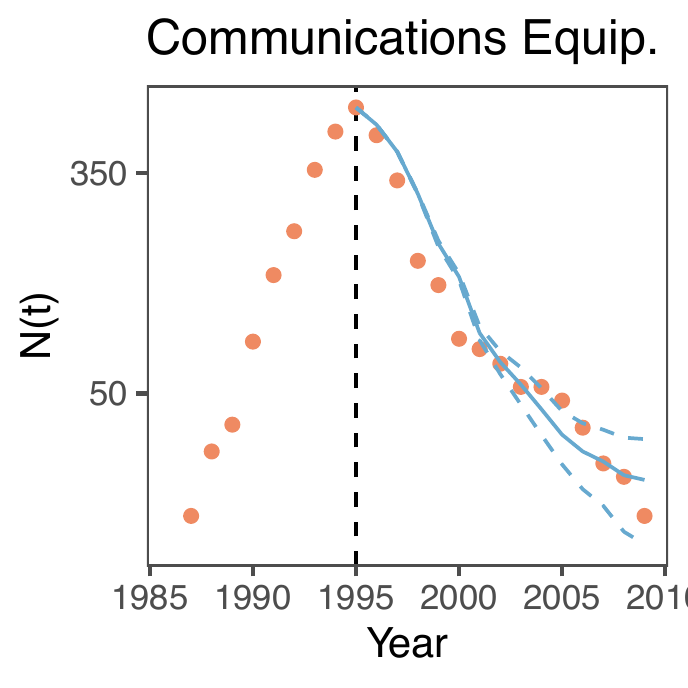}\hfill
\includegraphics[width=.3\textwidth]{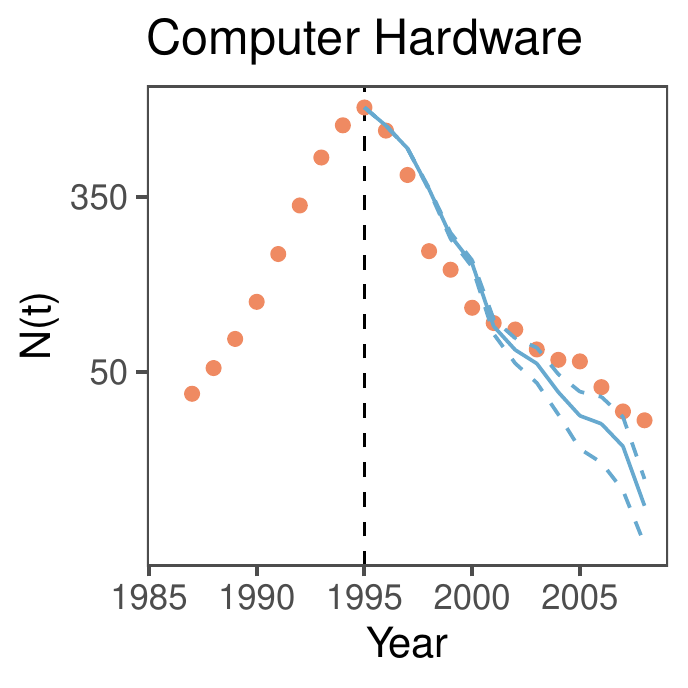}\hfill
\includegraphics[width=.3\textwidth]{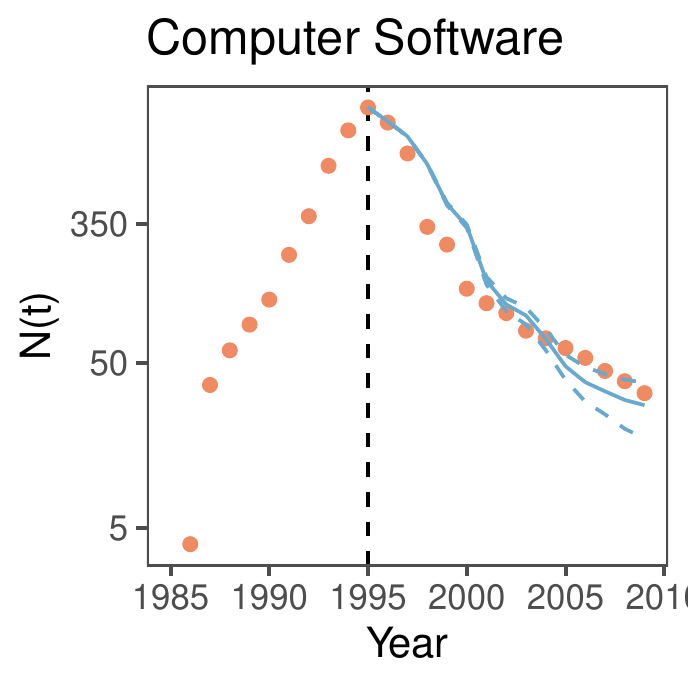}\\
\includegraphics[width=.3\textwidth]{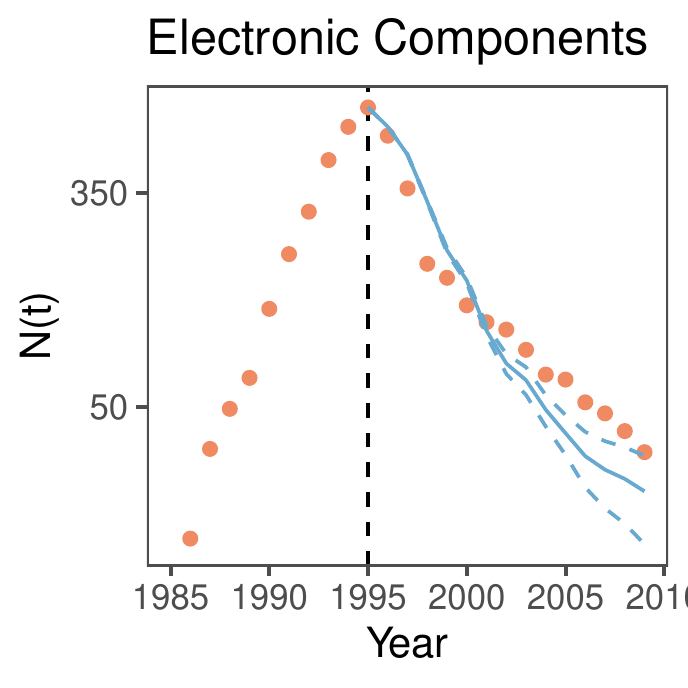}\hfill
\includegraphics[width=.3\textwidth]{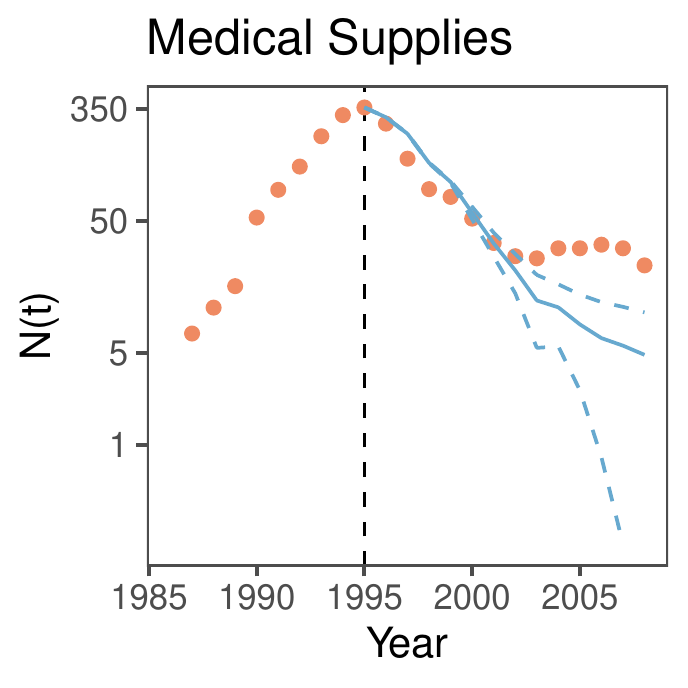}\hfill
\includegraphics[width=.3\textwidth]{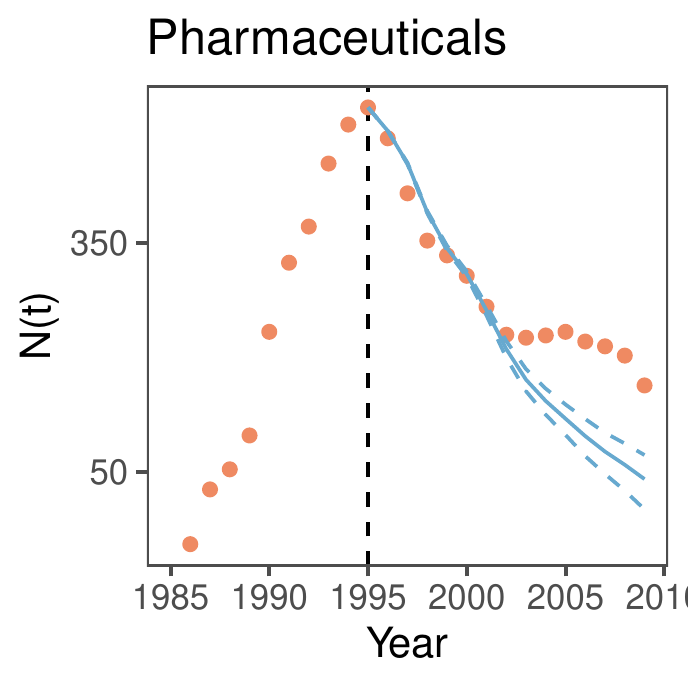}
\caption[]{Number of firms from the class of 95 remaining in the respective collaboration network after 1995.
  (yellow marks): empirical observations, (blue lines): predicted network size with 95\% confidence intervals (dashed lines). Note the $\log$ scale of the $y$ axis. 
  }
\label{fig:decline}
\end{figure}

Because we lack information about leaving dates, we have to generate our empirical observations for the class of 95 from data about their new alliances and about the life time of their established alliances, as before. 
Firms that are no longer part of any active alliance in a given year $t$ are assumed to leave.
This way, we obtain reference data about $N(t)$ that are plotted in Figure~\ref{fig:decline} (yellow marks).

Now we have to compare these data with our results from \emph{simulating the cascades}.
Instead of the information about active alliances, we now consider the probabilities of firms to leave the network.
The estimated leaving probabilities $p$ for the different sectors shown in Figure~\ref{fig:leaving_probabilities} denote a \emph{lower bound} because they were obtained considering the growth phase of the network.
To simulate the decline of the collaboration network, we need to adjust them over time, i.e., $p_{i}(t)$.
For 1995, this probability is given in the plots shown in Figure~\ref{fig:leaving_probabilities}. 
For each year $t$ \emph{after} 1995, i.e., from 1996 to 2009,
we then \emph{recompute} the probability that a firm which is present in year $t-1$ leaves in the coming year $t$.
For this recalculation, we take the information about the network at time $t-1$ into account, in particular about the number of active partners.
I.e., we recompute the plots shown in Figure~\ref{fig:leaving_probabilities} for every year $t$:
\begin{align}
  \label{eq:1}
  \ln\left[\frac{p(t)}{1-p(t)}\right]=\alpha+\beta N^{a}(t-1)
\end{align}
We note that with this incremental update we are far from just fitting the leaving probabilities to a given year.
Instead, we consider the history of the collaboration network, as well as the actual situation for firms regarding their active partners.

\begin{figure}[htbp]
\centering
\includegraphics[width=.3\textwidth]{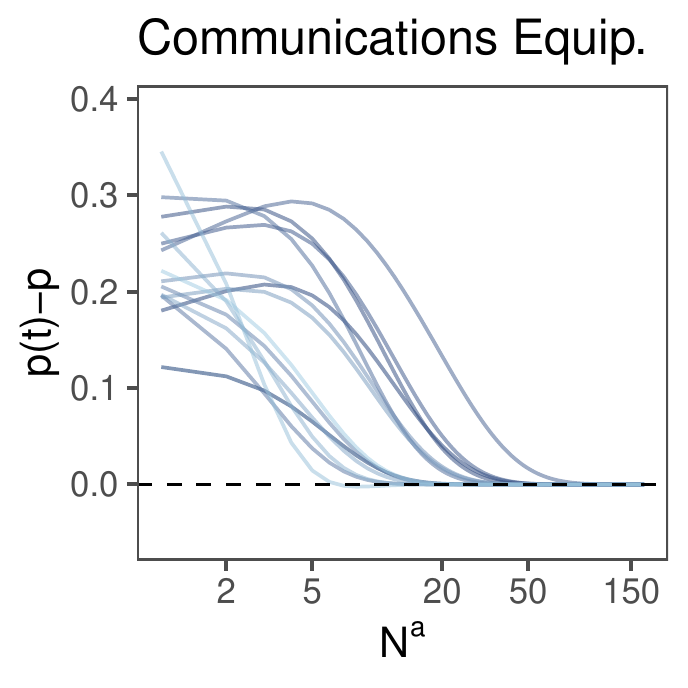}\hfill
\includegraphics[width=.3\textwidth]{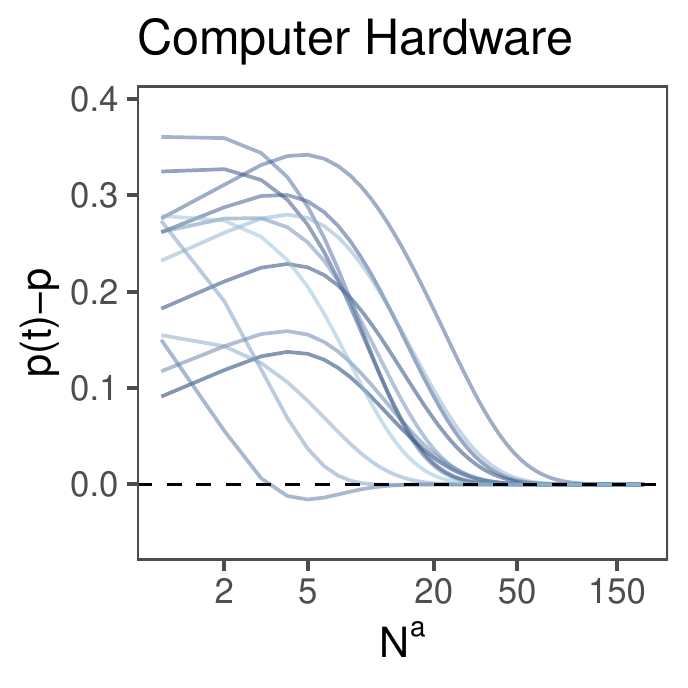}\hfill
\includegraphics[width=.3\textwidth]{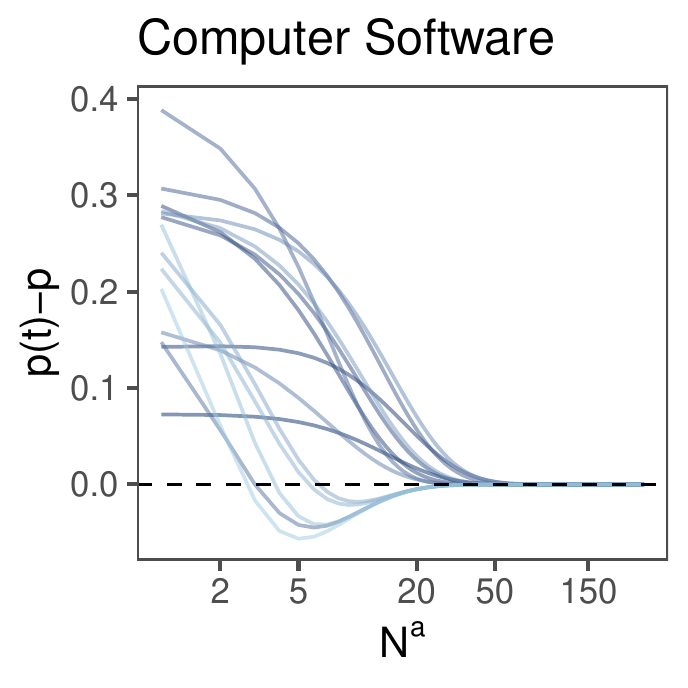}\\
\includegraphics[width=.3\textwidth]{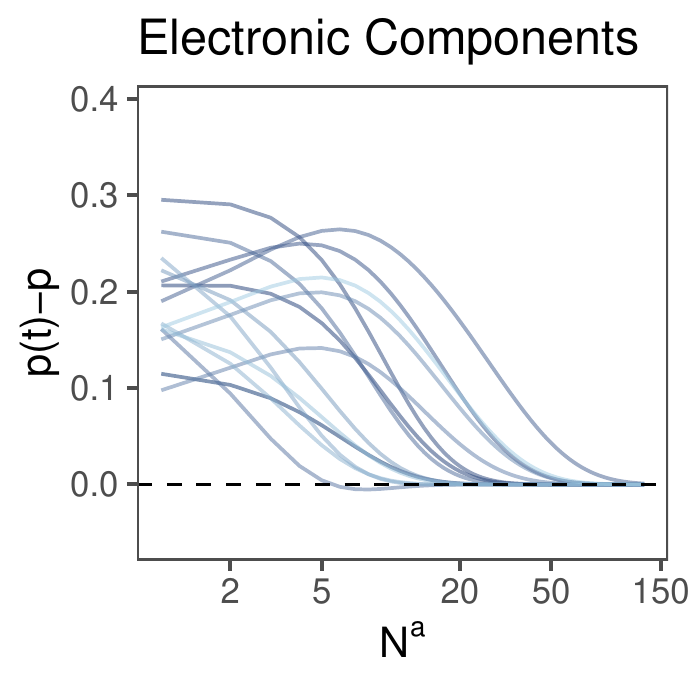}\hfill
\includegraphics[width=.3\textwidth]{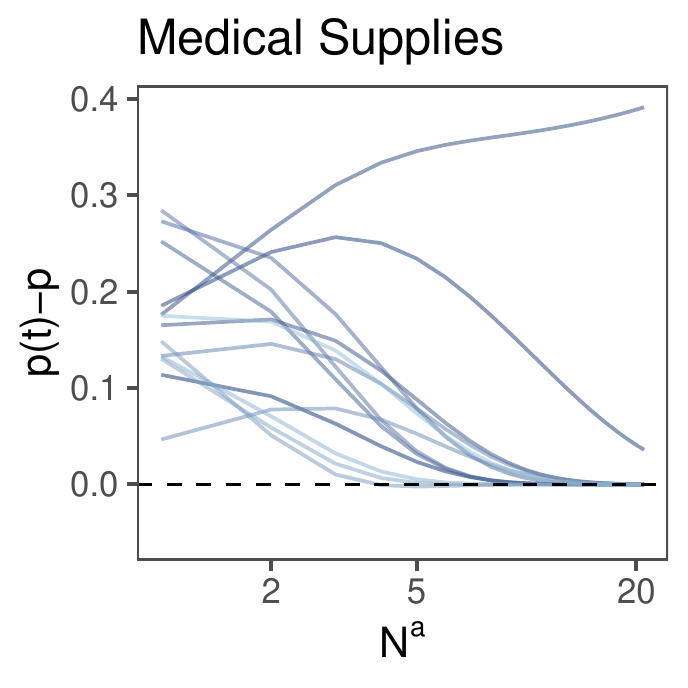}\hfill
\includegraphics[width=.3\textwidth]{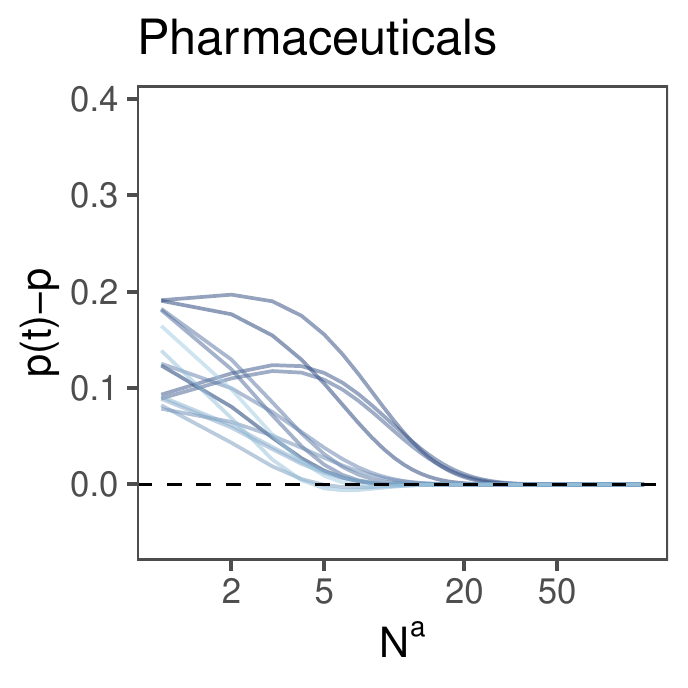}\\
\includegraphics[width=\textwidth]{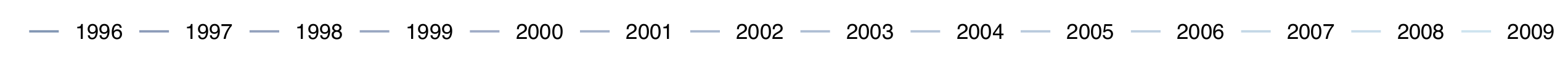}
\caption[]{Time dependent probability $p(t)-p$ for firms to leave the network dependent on the number of active partners. The blue scale encodes the years 1996-2009 with decreasing darkness. 
}
\label{fig:p-t}
\end{figure}

The regression results are shown in Figure~\ref{fig:p-t} for all industrial sectors.
For all years after 1995 we have plotted the differences $p(t)-p$, where $p$ refers to the values of 1995, as shown in Figure~\ref{fig:leaving_probabilities}. 

From the results, we notice that firms with a \emph{fewer} active partners are more affected by the time-dependent leaving probabilities than firms with many partners.
Specifically, for firms with less than 5 active partners the constant leaving probability underestimates their chances to leave, i.e., in reality they have left more often.
Only for two sectors, \emph{Computer Hardware} and \emph{Computer Software}, in few years
firms with 5 or more active partners leave \emph{less often}, i.e., the constant leaving probability overestimates their chances to leave.

With these adjusted probabilities, we make a prediction about the network in year $t+1$.
Precisely, we calculate the expected number $L(t)$ of firms from the class of 95 that will leave.
These firms are then removed together with their links and the collaboration network of the class of 95 is updated: $N(t+1)=N(t)-L(t)$. 
This way we obtain each year a small cascade of firms leaving, which sum up to the considerable decline of the network.
Our prediction for $N(t+1)$  is plotted in Figure~\ref{fig:decline} as the blue curve.
Because our simulations involve a stochastic component regarding the time when firms leave,
we have averaged this cascade dynamics over 200 runs.

Figure~\ref{fig:decline} shows both the empirical and the simulated network sizes for the six different industrial sectors.
We want to point out two observations.
First, it is remarkable how well our simulations of the network decay match the empirical network sizes for 4 out of 6 industrial sectors.
We note that this holds irrespective of the different industrial sectors and  the different size of the networks.
Arguably, \emph{Computer Software} and \emph{Electronic Components} refer to very different industries and to larger or smaller collaboration networks.
So, the agreement found lends evidence to the conclusion that the cascade dynamics we assumed indeed captures an essential mechanism of the observed decline. 

Second, it is as interesting to note the two cases where the simulated cascade dynamics does \emph{not} match the empirical decline: in \emph{Pharmaceuticals} and \emph{Medical Supplies} we clearly distinguish two phases of the decline.
In an earlier phase, from 1995 to 2000, our simulations still agree with the empirical sizes.
But in the last phase, from 2000 to 2005, they significantly deviate from the real evolution.
Our model would predict that the cascades are further amplified and even more firms from the class of 95 have left, whereas the empirical dynamics shows a remarkable stabilization.
The trend towards decline is stopped, instead the network size of the class of 95 remains almost constant until the end of the observation period.

This second observation motivates the discussion in the subsequent sections.
In a first step, we want to analyze how to improve the estimates for the leaving probabilities, to better reproduce the observed network sizes for the two cases of 
\emph{Pharmaceuticals} and \emph{Medical Supplies}.
In a second step, we discuss what determines these improved leaving probabilities.

\subsection{Adaptive leaving probability}
\label{sec:adapt-leav-prob}

\begin{figure}[htbp]
\centering
\includegraphics[width=.3\textwidth]{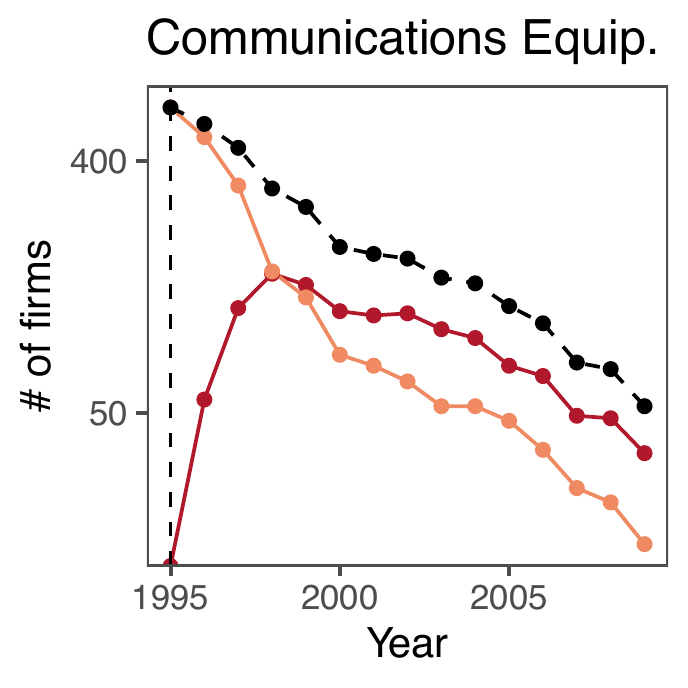}\hfill
\includegraphics[width=.3\textwidth]{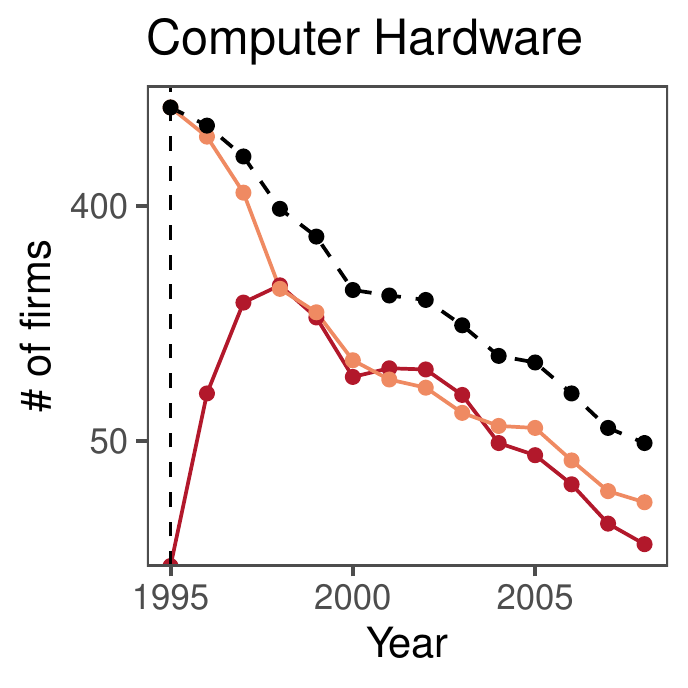}\hfill
\includegraphics[width=.3\textwidth]{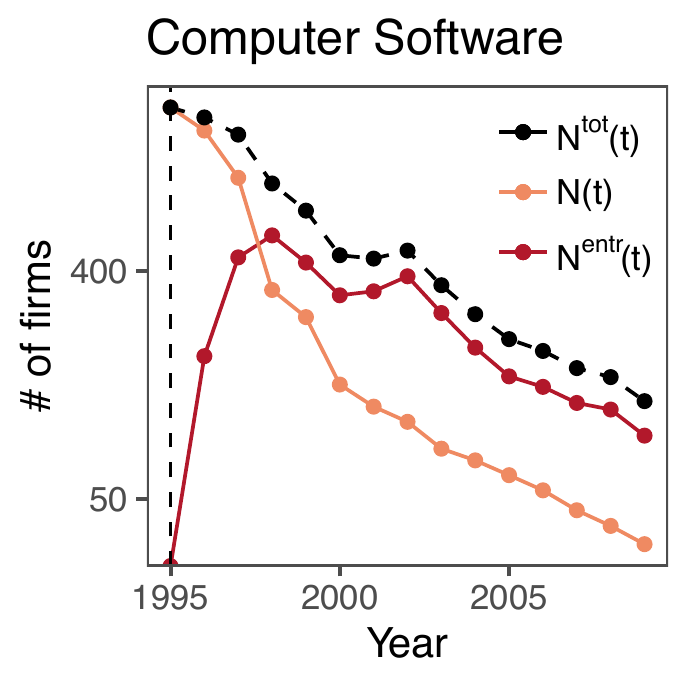}\\
\includegraphics[width=.3\textwidth]{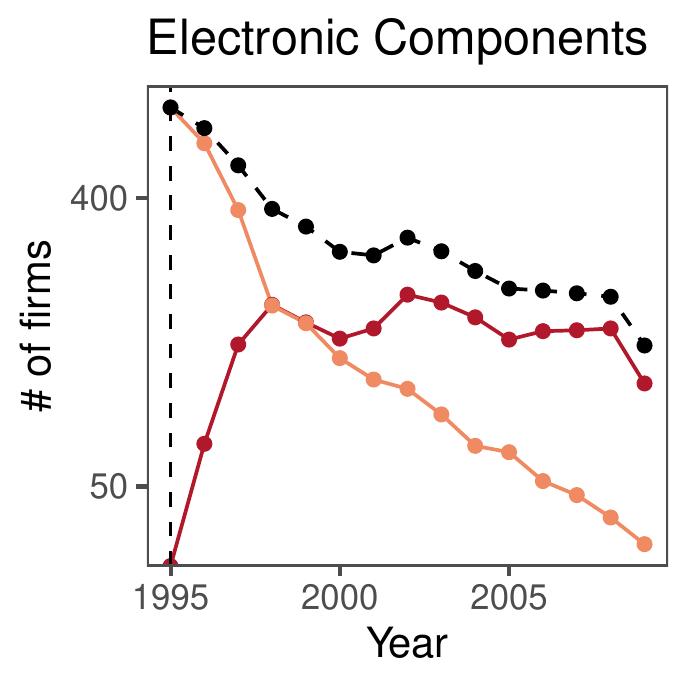}\hfill
\includegraphics[width=.3\textwidth]{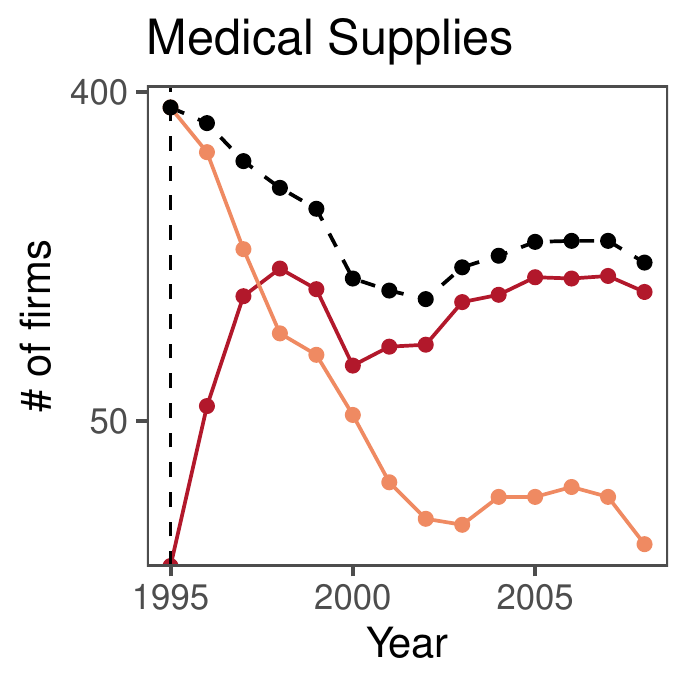}\hfill
\includegraphics[width=.3\textwidth]{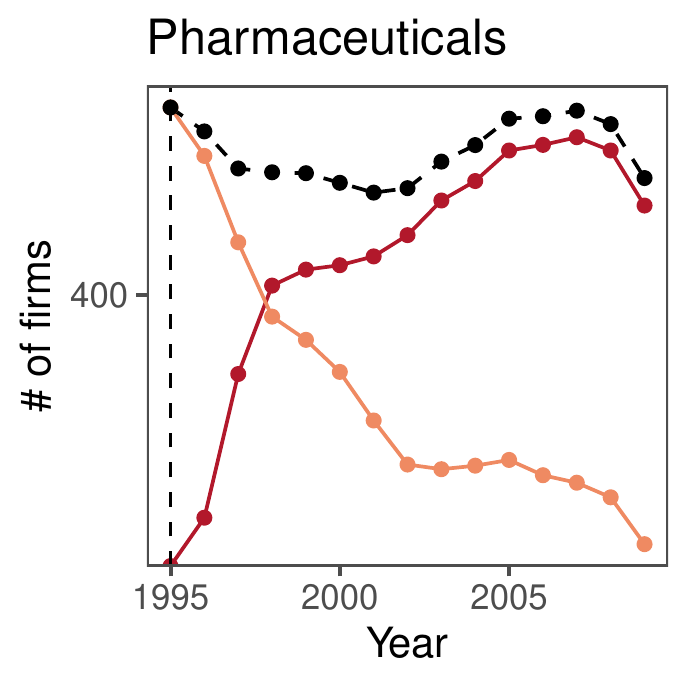}
\caption[]{Sizes of the collaboration networks  
  for different industrial sectors over time. Shown is the total network size $N^{\mathrm{tot}}(t)$, the number of firms from the class of 95, $N(t)$ and the number of newcomers, $N^{\mathrm{entr}}(t)$.
  Note the log scale of the $y$ axis.
}
\label{fig:N3t}
\end{figure}

What additional information do we have available to further improve the estimates for the leaving probabilities? 
So far, we have used only information from firms of the class of 95.
But the collaboration network changes not only because of the \emph{exit} of the established firms from the class of 95, there is also the \emph{entry} of new firms.
Despite the overall ``rise and fall'' trend, where decline dominates after 1995, a considerable number of newcomers enter the existing networks each year.
Figure~\ref{fig:N3t} shows the total number of firms in the network, $N^{\mathrm{tot}}(t)$, the number of firms remaining from the class of 95, $N(t)$, and the number of \emph{new firms} entering the network in each year after 1995, $N^{\mathtt{entr}}(t)$.
By construction, in our reference year 1995 $N=N^{\mathrm{tot}}$, $N^{\mathrm{entr}}=0$.

As we observe, in most sectors the total network size $N^{\mathrm{tot}}(t)$ declines over time, even though we have the entry of new firms.
So this does not break the overall trend. 
The exceptions are the two sectors \emph{Medical supplies} and \emph{Pharmaceuticals}, and in these we are mostly interested. 
The difference results from two combined effects, as Figure~\ref{fig:N3t} indicates: 
(a) the very pronounced increase in the number of newcomers, and (b) the slowed down decrease in the number of firms from the class of 95.
We see that in all sectors already from 1998 on the number of newcomers \emph{exceeds} the number of those firms from the class of 95 that still remained in the network.
But for the two sectors \emph{Medical supplies} and \emph{Pharmaceuticals} the year  1998 is the time when the number of firms from the class of 95 stops plunging and exhibits a slower decline instead.
The most suitable interpretation for this observation is in fact that these newcomers basically prevent the established firms from leaving the network.

We will discuss this argument in more detail in Section~\ref{sec:discussion}.
Before, we want to check whether information about the newcomers would allow us to improve the estimates about the leaving probabilities. 
We repeat the procedure to calculate $p(t)$, Eqn.~(\ref{eq:1}), but now we correct the values for $N^{a}(t)$ to take the newcomers into account.
Then we repeat the simulations of the cascades shown in Figure~\ref{fig:decline} with the adaptive leaving probabilities.

\begin{figure}[htbp]
\centering
\includegraphics[width=.3\textwidth]{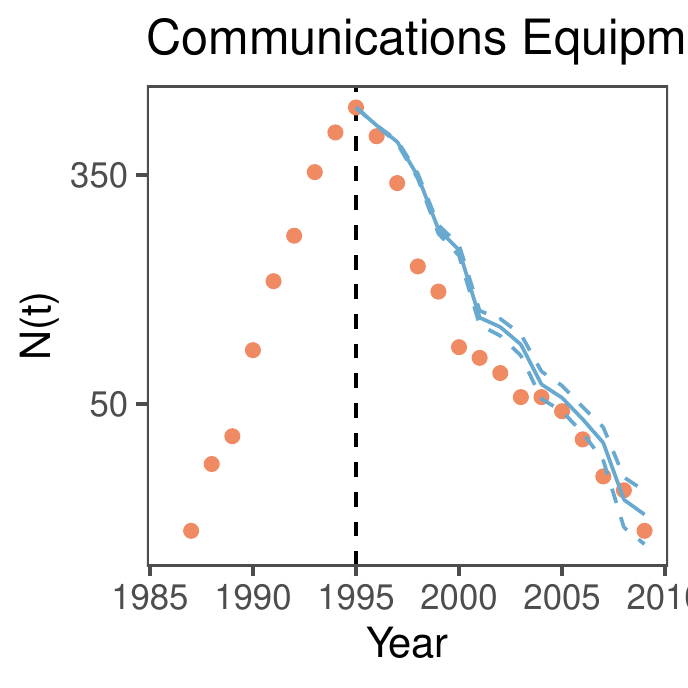}\hfill
\includegraphics[width=.3\textwidth]{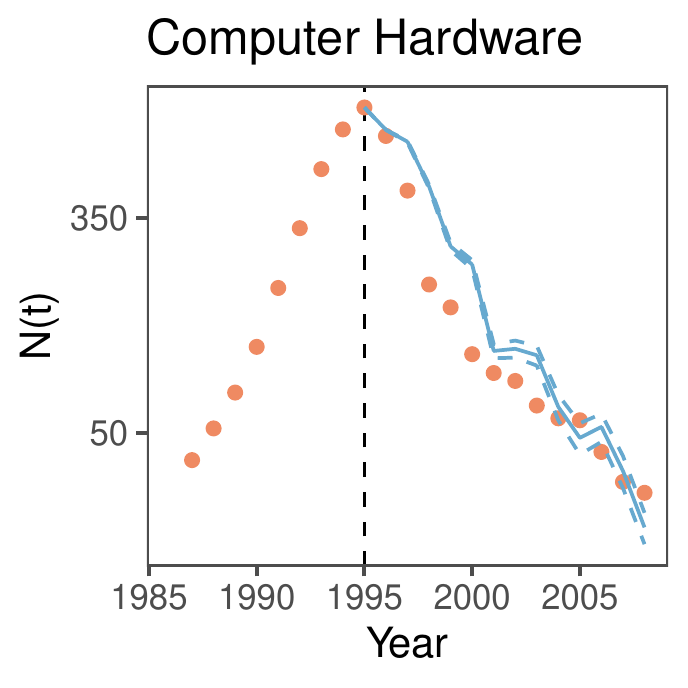}\hfill
\includegraphics[width=.3\textwidth]{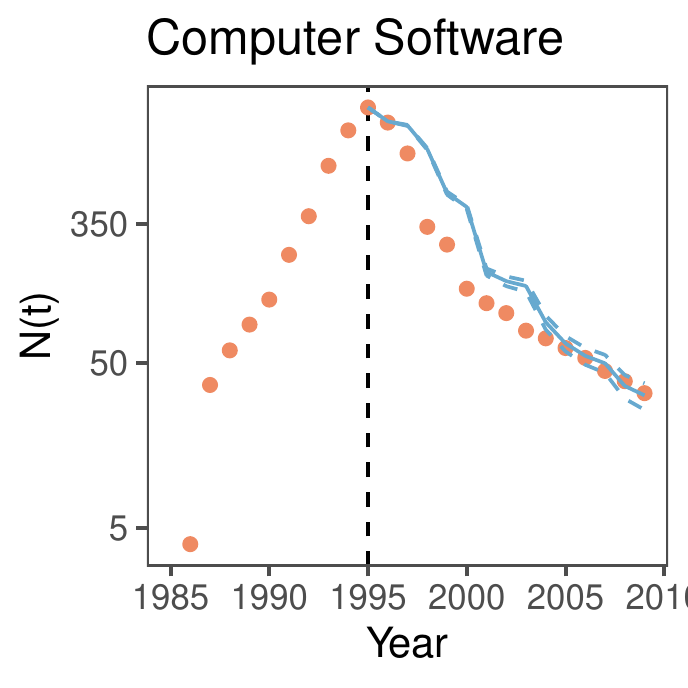}\\
\includegraphics[width=.3\textwidth]{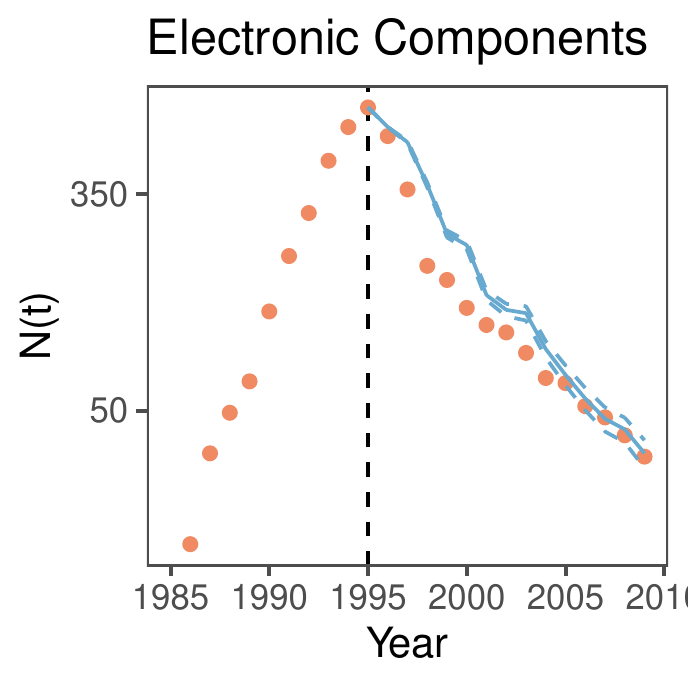}\hfill
\includegraphics[width=.3\textwidth]{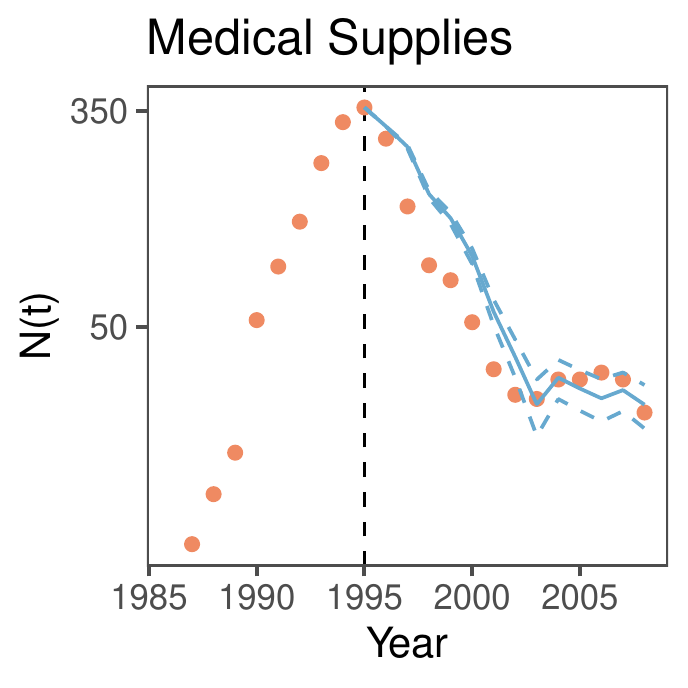}\hfill
\includegraphics[width=.3\textwidth]{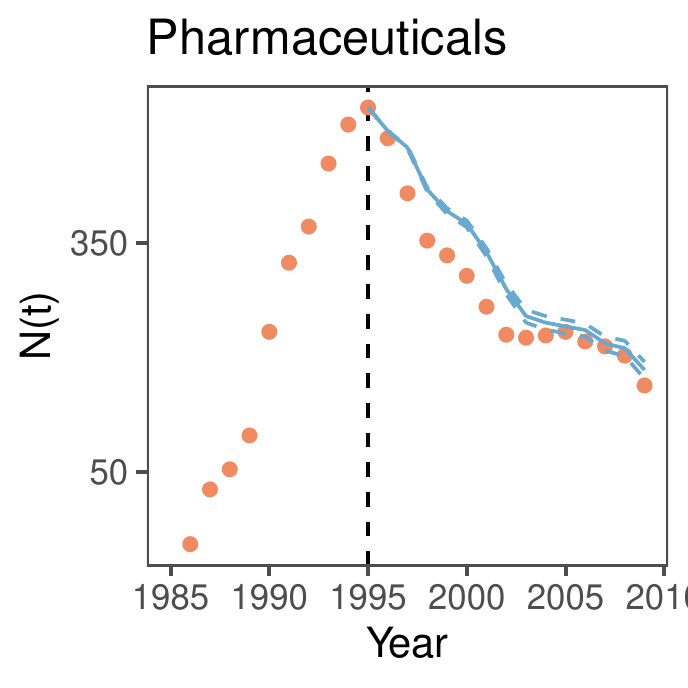}

\caption[]{Number of firms from the class of 95 remaining in the respective collaboration network after 1995.
  (yellow marks): empirical observations, (blue lines): predicted network size with 95\% confidence intervals (dashed lines). The model has used the adaptive leaving probabilities, obtained from Eqn.~\eqref{eq:1} corrected for the newcomers. 
}
\label{fig:temporal_indicator}
\end{figure}

The results are shown in Figure~\ref{fig:temporal_indicator}.
They demonstrate that with the adaptive leaving probabilities we can accurately model the network decline of the firms from the  class of 95 for each year.
This now holds for all industrial sectors, even for \emph{Pharmaceuticals} and \emph{Medical Supplies}.
Thus, our cascade model with adaptive leaving probabilities that takes the impact of newcomers into account is able to reproduce the empirical network sizes.

\section{Discussion}
\label{sec:discussion}

\subsection{Improved adaptivity}
\label{sec:improved-adaptivity}

The very good agreement between the empirical and simulated network sizes shown in Figure~\ref{fig:temporal_indicator} lends evidence to our methodology to estimate the leaving probabilities of firms.
In particular, it supports our underlying assumption, Eqn.~(\ref{eq:1}), that the number of active partners is a main constituent for their decision to stay or to leave the network.
But we learned that we have to correct this number to accommodate for the entry of new firms, to obtain the good results. 

\begin{figure}[htbp]
  \centering
  \includegraphics[width=.45\textwidth]{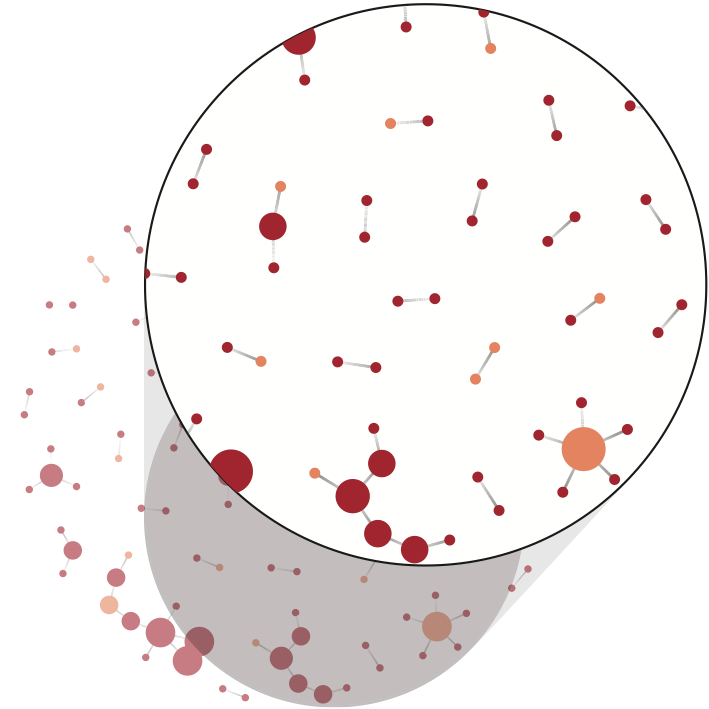}
  \includegraphics[width=.45\textwidth]{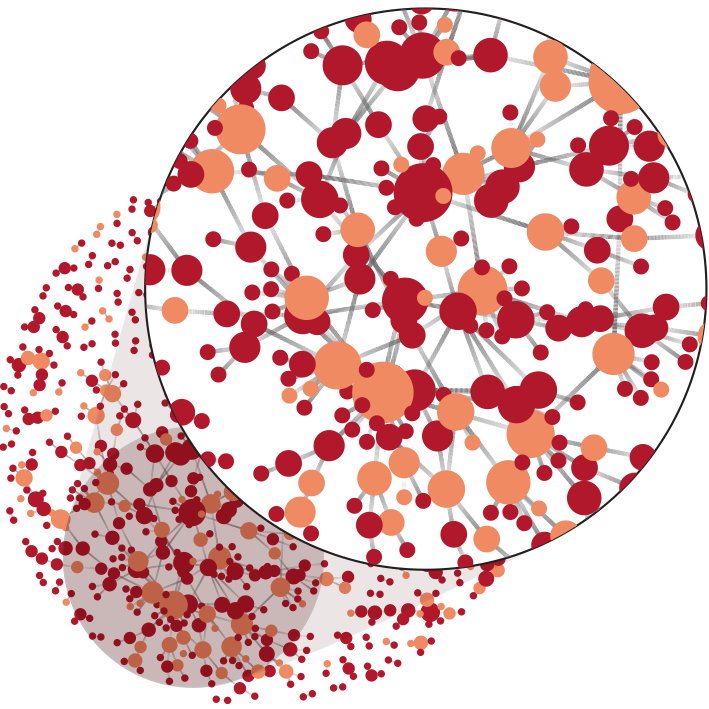}
  \caption{Snapshots of the collaboration networks in 2003: (left) \emph{Medical Supplies}, (right) \emph{Pharmaceuticals}. (orange) firms of the class of 95, (red) newcomers. 
}
  \label{fig:adapt}
\end{figure}

This leaves us with the task to explain \emph{why} in some cases the newcomers have such a remarkable influence.
As already mentioned, we argue that these newcomers prevent the established firms from leaving the network because
they provide new opportunities to collaborate and often also bring innovative knowledge to the collaboration network (start-ups).
Instead of relying on the collaboration with established firms, the firms from the class of 95 now adapt to the situation.
They form new R\&D alliances with the newcomers, increasing their benefits and have no further reason to leave.
Figure~\ref{fig:adapt} shows two snapshots from the collaboration networks in \emph{Medical Supplies} and \emph{Pharmaceuticals}, to illustrate this interpretation.
These insights can be turned into a novel argument about the \emph{resilience} of networks.
What we observe is the \emph{adaptive ability}, or adaptivity, of established firms to cope with the new situation.
Instead of following the trend to leave the network, they find new ways of leveraging the situation.
This dynamics is precisely what the term \emph{resilience} shall describe: the capacity to withstand shocks generated by the leave of active partners, and the ability to recover from these shocks, by establishing relations to new partners.
What sounds reasonable for a personal life (and has inspired early definitions of resilience in a psychological context), can be observed also for firms, as our analysis reveals.

\subsection{Fragile, yet resilient}
\label{sec:frag-yet-resil}

It is the reason for the time-dependent change of the leaving probabilities, $p(t)$, that the collaboration networks have \emph{adapted} to the situation of continuous decline.
We note that this decline has not completely stopped.
Compared to the golden age of 1995, all networks have become much more \emph{fragile}.
Many firms have left, established collaborations ceased to exist.
But some networks are still \emph{resilient} in the sense described above.
Those firms that managed to stay in the network after the ``fall'' trend took over, are indeed the seed for this resilience.
They offer newcomers possibilities to integrate into the, this time much smaller, collaboration network and they ``connect the dots'', as the backbone of the network.
Thus, the decline of the network has offered the chance, more correctly it increased the pressure, for the network to adapt to a changing environment of R\&D collaborations.

This leaves us with the question whether our findings could simply be reduced to the fact that newcomers enter the network.
This assumes that a high entry rate would be sufficient to make a network resilient.
We can refute this argument with reference to an earlier study about the ``autopsy'' of the social network \emph{Friendster} \citep{mavrodiev2013}.
This network collapsed despite a size of 113 million users.
New users always entered the network until the very end.
But it was shown that after the network has reached a size of 80 million users, the more than 30 million new users still entering became less integrated into the network.
Hence, what matters is not the network growth, i.e., the rate at which new nodes enter the network.
Whether or not the network becomes resilient depends on the \emph{intergration} of these new nodes into the network.
\emph{Friendster} failed in this respect and collapsed despite a steady growth.

As the two snapshots of Figure~\ref{fig:adapt} and the dynamics in Figure~\ref{fig:decline} show, the R\&D collaboration networks for \emph{Medical Supplies} and \emph{Pharmaceuticals} were successful in integrating newcomers.
That's why the established firms continued to stay.
This does not mean that the network has to be as dense as for \emph{Pharmaceuticals}.
As we have already shown in Figure~\ref{fig:leaving_probabilities}, each sector is characterized by a different cost-benefit relation which determines the conditions for firms to leave.
In case of \emph{Medical Supplies}, it is obviously sufficient that established firms start collaborating with 1-2 newcomers, whereas for \emph{Pharmaceuticals} the critical number of active partners has to be higher.

To conclude, after 1995 all collaboration networks have become \emph{fragile}, indicated by the global decline trend.
To some degree,they are \emph{yet resilient} dependent on their ability to \emph{integrate newcomers}.
The phrase ``fragile, yet resilient'' makes reference to an early study about the robustness of infrastructure networks, such as the internet, which were dubbed as ``robust, yet fragile'' \cite{Doyle14497}. 
There the term ``fragile'' referred to the fact that networks with a very broad degree distribution are vulnerable against the removal of nodes with a high degree.
Such nodes are rare, therefore a random removal of nodes would most likely hit one of the many nodes with a very low degree.
But a targeted attack, if focused on the high-degree nodes, can easily destroy the network.
This insight, however, refers to the expected properties of an ensemble of scale-free networks and cannot be applied to all individual realizations.
The internet, in particular, has a low probability to occur at random.
It is  carefully designed for robustness and therefore much less fragile than random realizations.

A similar discussion also applies here.
On the one hand, we observe cascades of firms leaving the network because they have less active partners, which in turn increases the trend.
This denotes the expected behavior of a network break-down.
The double feedback that amplifies this cascade, namely that over time more and more firms have less and less active partners for collaborations, is also known from other cascade models, e.g., from the so-called fiber bundle model \cite{kun2000damage,Lorenz2009}.
On the other hand, because collaboration networks are \emph{adaptive}, they have in principle the ability to deviate from this expected behavior.
Even more, we can turn this deviation from the expected behavior into a measures of the \emph{adaptivity} of the system and, because it prevents the breakdown, as a measure of \emph{resilience}.

As our results illustrate, not all collaboration networks in the different industrial sectors show this adaptive behavior to the same degree. 
Hence, their decline has continued as expected.
We can only speculate why the networks in \emph{Medical Supplies} and \emph{Pharmaceuticals} seem to be more adaptive and thus more resilient.
Two arguments come into play.
One refers to the large number of newcomers which offer ample new opportunities.
Economically, this points to low barriers for firms to enter the market, but also to an increased dependency of the industry on external innovations.
In \emph{Pharmaceuticals}, for instance, start-up firms provide a large share of new technologies, substances, etc.
The second argument, however, is as important, namely the ability of established firms to integrate these newcomers into their own R\&D activities.
This largely depends on legal constraints, such as compliance or protection of intellectual properties, but also on the economic pressure to exploit innovative knowledge earlier than the competitors.
Indeed, empirical evidence \cite{EPJDS-Cross-2017a} shows that in the sector \emph{Pharmaceuticals}  established firms have a higher probability to form alliances with newcomers (30\% of newly formed alliances) than in all other studied sectors, which exhibited probabilities ranging from 10\% (\emph{Communications Equipment}) to 25\% (\emph{Computer Software}). 

With this discussion we have provided an interface toward economics, in a truly interdisciplinary manner, which can be explored in the future. 
But the research presented in this paper also offers a general insight for network science, where studies about declining networks are still rare.
Obsessed with network growth and stability, one should try to avoid the premature focus on the general trend.
Decline is not a synonym for instability and a precursor of collapse. 
As often it is part of a life-cycle dynamics, where decline should be expected rather than feared.
As we have demonstrated, firms, as individuals in a social setting, have the ability to cope with this trend, this way making the system more resilient than expected.
Hence, quantitative measures for resilient networks cannot be simply taken from the evolution of the network size.
It needs a deeper reflection about the problem of resilience in face of a life cycle, which we just started to provide here.

\subsection*{Acknowledgement}
\label{sec:acknowledgement}

This research received financial support from the ETH Foundation through the ETH Risk Center Seed Projects
``Performance and Resilience of Collaboration Networks'' (MT) and ``Systemic Risks for Privacy in Online Interaction'' (DG).

\small \setlength{\bibsep}{1pt}

\end{document}